\documentclass[twocolumn,showpacs]{revtex4}
\usepackage[xdvi]{graphicx}%
\usepackage{dcolumn}
\usepackage{amsmath}
\makeatletter

\def\btt#1{\texttt{\@backslashchar#1}}%
\DeclareRobustCommand\bblash{\btt{\@backslashchar}}%
\makeatother

\newcommand{\bra}{\left\langle}
\newcommand{\ket}{\right\rangle}
\newcommand{\brap}{\left.}
\newcommand{\ketp}{\right.}
\newcommand{\ketc}{\right\rangle_{\rm c}}

\newcommand{\der}[2]{\frac{d #1}{d  #2}}

\newcommand{\vs}{v_{\rm s}} 
 
\newcommand{\fp}{f^{\rm p}} 
\newcommand{\Qp}{Q^{\rm p}} 
\newcommand{\Wp}{W^{\rm p}} 
 \newcommand{\mud}{\mu_{\rm d}}

\newcommand{\cD}{{\cal D}}
\newcommand{\cA}{{\cal A}}
\newcommand{\e}{{\rm e}}

\begin{document}

\title{Decomposition of force fluctuations far from equilibrium}
\author{Kumiko Hayashi and Shin-ichi Sasa}
\affiliation
%\address
{Department of Pure and Applied Sciences,  
University of Tokyo, Komaba, Tokyo 153-8902, Japan}

\date{\today}

\begin{abstract} 
By studying a nonequilibrium Langevin system, we find that a
simple condition determines the decomposition of the coarse-grained 
force into a dissipative force, an effective driving force 
and noise.  From  this condition,  we derive a new universal 
inequality,  $D \ge \gamma \mud^2 T$, relating the diffusion 
constant $D$, the differential mobility $\mud$, the bare friction 
constant $\gamma$ and the temperature $T$. Due to the general nature 
of the argument we present, we believe that our idea concerning this 
decomposition can be applied to a wide class of systems far from 
equilibrium.
\end{abstract}

\pacs{05.40.-a, 02.50.Ey, 87.16.Nn}
\maketitle

%%%%%%  introduction  %%%%%

%%% general 

The nature of a force depends on the scale on which it is observed. 
For example, consider a force exerted by water molecules on a 
colloidal particle. Such a force can be described by mechanical 
laws on  molecular time scales, while it is described as a 
dissipative (frictional) force and thermal noise on time scales 
of the order of $10^{-3}$ sec. In analogy to this example, for a 
wide range of systems, including bio-mechanical systems \cite{ex1} 
and  granular systems \cite{ex2.1}, it might be expected 
that a fluctuating force obtained through  some  coarse-graining  
procedure  can  be decomposed into a dissipative force and  other 
components.  In the regime near equilibrium, such a decomposition 
is uniquely determined by fluctuation-dissipation relations (FDRs). 
However, no rule is known that determines such a decomposition far 
from equilibrium \cite{fn:proj}.  We wish to discover a rule of 
this kind for a class of systems exhibiting fluctuating forces. 

%%% model 

With the above-stated purpose, in the present work, we study a 
Langevin equation describing the motion of a Brownian particle 
with a tilted periodic potential in a one-dimensional space. 
Although we study the simplest system realizing nonequilibrium 
steady states (NESSs), the arguments below can be applied to a 
wide class of Brownian motors  
\cite{Butt,fla,pul}. The Langevin equation that 
we analyze is described by  
\begin{equation}
\gamma\der{x}{t}=f-\der{ U(x)}{x}+\xi(t).
\label{model1} 
\end{equation}
Here, $\gamma$ is a friction constant,  
$U(x)$ is a periodic potential of period $\ell$, $f$ is a 
constant external driving force, and $\xi$ is  Gaussian white 
noise satisfying   
\begin{equation}
\bra\xi(t)\xi(t^{'})\ket=2\gamma T\delta(t-t^{'}),
\label{noise} 
\end{equation}
where $T$ is the temperature of the environment, and the Boltzmann 
constant is set to unity.  We consider a description of large-scale 
motion,  which is obtained by taking an average over a time interval 
$\delta t$ that is chosen to be sufficiently longer than the  
characteristic time of the system. In this description, the finite 
time average of the force $-dU/dx$ acts on the particle as a 
fluctuating force. We conjecture that this fluctuating force can be 
decomposed into a dissipative force, an   extra driving force and 
random noise.  

%%% time average 

In order to investigate time averaged quantities, including that 
of $-dU/dx$,  we introduce the finite time average of an arbitrary 
quantity $Z(t)$: 
\begin{equation}
\overline{Z}_n\equiv \frac{1}{\delta t}\int_{t_n}^{t_{n+1}} dt  Z(t), 
\end{equation}
where $t_n=n \delta t$, $n=0,1,2 \cdots$.  Then, the finite time 
average of $-d U/d x$ that we consider is given by 
\begin{equation}
-\overline{\der{U}{x}}_n = -\frac{1}{\delta t}\int_{t_n}^{t_{n+1}} 
dt\left. \der{U}{x} \right\vert_{x=x(t)}.  
\end{equation}
We hypothesize that this can be decomposed into a dissipative 
component $A (x_{n+1}-x_n)/\delta t$, where $x_n\equiv x(t_n)$, and  
a non-dissipative component.   That is, we assume the form
\begin{equation}
-\overline{\der{U}{x}}_n=A \frac{x_{n+1}-x_n}{\delta t}+B_n,
\label{decom}
\end{equation}
where $A$ is a constant and $B_n$ is a fluctuating quantity whose 
statistical average $\bra B_n\ket$ takes a non-zero value $\fp$: 
\begin{equation}
B_n=\fp+\delta B_n. 
\end{equation}
The quantities $\fp$  and $\delta B_n$ correspond to an extra 
driving force and 
random noise, respectively.  Substituting (\ref{decom}) into an 
integrated form of (\ref{model1}),  we obtain 
\begin{equation}
(\gamma-A) \frac{x_{n+1}-x_n}{\delta t}=f+ \fp +\delta B_n
+\overline{\xi}_n.
\label{model3}
\end{equation}
Then, with the definitions
\begin{eqnarray}
\Gamma &\equiv& \gamma-A,
\label{gamma} \\
 F &\equiv& f+\fp ,
\label{Force} \\
 \Xi_n &\equiv& \delta B_n+\overline{\xi}_n , 
\label{Noise}
\end{eqnarray}
we express (\ref{model3}) as
\begin{equation}
\Gamma \frac{x_{n+1}-x_n}{\delta t}=F +\Xi_n,
\label{model2}
\end{equation} 
where $\Xi_n$ is expected to exhibit a Gaussian distribution for 
large $\delta t$. This equation is regarded as an effective model 
of (\ref{model1}). 

The main claim  of the present Letter is that the simple condition
\begin{equation}
\lim_{\delta t\to\infty} \delta t \bra \delta B_n \overline\xi_n \ket =0
\label{hcon}
\end{equation}
uniquely determines the constant $A$ in (\ref{decom}). Note that 
a correlation of time averaged quantities is proportional to 
$\delta t^{-1}$ in general and (\ref{hcon}) indicates that the 
proportional constant for the case $\bra \delta B_n \overline\xi_n 
\ket$ becomes zero. 
The condition (\ref{hcon}) implies that $A (x_{n+1}-x_n)/\delta t$,  
which fluctuates in time, can be distinguished from $B_n$ by the 
condition $(\ref{hcon})$ \cite{fn:decom}. After presenting the proof 
of this claim, we remark on three important topics related to it: 
a general inequality obtained as a direct application of (\ref{hcon}), 
energetic considerations related to (\ref{hcon}), and the relation 
between (\ref{hcon}) and  a time reversal symmetry in the stochastic 
sense. 

%%%%%%  end of  introduction  %%%%%

%%%%%%%%%%%%%%%%%%%%%%%%%%%%%%%%%%%%%%%
%%%%%%%%%%%%%%%%%%%%%%%%%%%%%%%%%%%%%%%
\paragraph*{Preliminary consideration:}
%%%%%%%%%%%%%%%%%%%%%%%%%%%%%%%%%%%%%%%
%%%%%%%%%%%%%%%%%%%%%%%%%%%%%%%%%%%%%%%

Before presenting  the proof of the main claim, we first consider 
how the parameters of the effective model (\ref{model2}) can be
expressed in terms of the steady state velocity $\vs$ and  the 
diffusion constant $D$, defined as  
$\vs \equiv  \lim_{t \to \infty} \bra (x(t)-x(0))/t \ket$ and 
$D \equiv  \lim_{t \to \infty} \bra (x(t)-x(0)-\vs t)^2/2t \ket$.  
Because $\vs$ and $D$ are independent of the scale on which we 
describe the system, the same values of $\vs$ and $D$ 
should be obtained from (\ref{model2}).  This implies the relations
$F=\Gamma\vs$ and 
\begin{equation}
\bra \Xi_n \Xi_m \ket =2 \delta_{nm} D\Gamma^2 (\delta t)^{-1}.
\label{noise:eff}
\end{equation}
From these expressions, all the parameters of the effective model 
(\ref{model2}) are given in terms of $\vs$ and $D$ when $\Gamma$ is 
determined. 

%% equilibrium case

In the equilibrium case ($f=0$), $\Gamma$ should satisfy the 
relation $D\Gamma^2=\Gamma T$, which is referred to as the FDR of 
the second kind \cite{kubo}. {}From this, $\Gamma$ is expressed as
\begin{equation}
\Gamma=\frac{T}{D}.
\label{2nd}
\end{equation}
Furthermore, we can prove 
\begin{equation}
D=\mu T,
\label{1st}
\end{equation}
where $\mu $ is the mobility, defined as  
$\mu= \lim_{f \to 0} \vs(f)/f$.  The two FDRs (\ref{2nd}) 
and (\ref{1st}) lead to 
\begin{equation}
\Gamma=\mu^{-1}.
\label{eqgam}
\end{equation}
However,  for NESSs far from equilibrium, (\ref{1st}) is violated 
\cite{HSIIIa}, and (\ref{2nd}) does not hold in general.
Therefore, for treatment of such systems, 
it is necessary to find a guiding principle to determine $\Gamma$. 

%% potential variation method in NESS

In a previous work \cite{HSIIIa}, we studied NESSs and proposed a 
natural method to determine $\Gamma$ by considering the response 
of the particle to  a slowly varying potential $V(x)$ in space.  
Below, we present a heuristic argument from which the result 
obtained there can be understood. For detailed presentation of 
the systematic perturbation method used to derive this result, 
see Ref. \cite{HSIIIa}. 

Because the gradient of the slowly varying potential $V(x)$ can be 
regarded as a modulation of the external force $f$, the large-scale 
motion in the modulated system is described by
\begin{eqnarray}
\frac{x_{n+1}-x_n}{\delta t}&\simeq& \vs \left(f-\frac{d V(x_n)}{dx_n}
\right)+\frac{\Xi_n}{\Gamma}   \nonumber \\
&\simeq& \vs(f) -\mud\frac{d V }{dx_n}+\frac{\Xi_n}{\Gamma}, 
\label{heu}
\end{eqnarray}
where $\mud$ is the differential mobility defined by 
\begin{equation}
\mud\equiv \frac{d \vs(f)}{df}. 
\label{mud}
\end{equation}
Then,  assuming that $-d V/dx_n$ is the force acting on the particle 
even in this effective description, from (\ref{heu}) we obtain the 
result 
\begin{equation}
\Gamma=\mud^{-1}.
\label{Gamma}
\end{equation}
Note that this expression represents an extension of (\ref{eqgam}) 
to the presently considered non-equilibrium case. 
{}From (\ref{gamma}) and (\ref{Gamma}), we find that the 
constant $A$ appearing  in  (\ref{decom}) should  satisfy 
\begin{equation}
\gamma-A=\mud^{-1}.
\label{gamA}
\end{equation}

%%%%%%%%%%%%%%%%%%%
%%%%%%%%%%%%%%%%%%%
\paragraph*{Proof:}
%%%%%%%%%%%%%%%%%%%
%%%%%%%%%%%%%%%%%%%

We demonstrate that  the decomposition condition (\ref{hcon}) 
uniquely determines the constant $A$, yielding  (\ref{gamA}).
%
%%%  proof 1 
%
Using (\ref{decom}) and (\ref{model3}), we rewrite the condition 
(\ref{hcon}) as 
\begin{eqnarray} 
& &\lim_{\delta t\to\infty} \delta t 
\bra \left(\overline{\der{U}{x}}_n + A \frac{x_{n+1}-x_n}{\delta t} 
\right) \ketp\nonumber \\   
& & \brap \left(\gamma \frac{x_{n+1}-x_n}{\delta t}
+\overline{\der{U}{x}}_n \right) \ketc =0,
\label{con2}
\end{eqnarray}
where $\bra \ \ketc$ represents the cumulant. Through the definition 
\begin{equation}
G \equiv -\lim_{\delta t\to\infty}\delta t \bra \overline{\der{U}{x}}_n 
 \cdot \frac{x_{n+1}-x_n}{\delta t} \ketc,  
\end{equation} 
it is easy to obtain  
\begin{equation}
\gamma-A= \frac{2\gamma T}{2\gamma D-G}.
\label{final}
\end{equation}

%%% proof 2

In order to connect (\ref{final})  with the 
differential mobility  $\mud$ defined by (\ref{mud}),  we express 
it in terms of $D$ and $\vs$.  We start with the path integral 
representation
\begin{eqnarray}
\bra  x(\delta t)-x(0) \ket &=& \int \cD x 
(x(\delta t)-x(0)) 
\nonumber \\
&\phantom{=}&\e^{-\frac{1}{4\gamma T}\int_{-\infty}^{\delta t} dt 
\left( \gamma \dot x-f+\der{U}{x} \right)^2}. 
\end{eqnarray}
Then, differentiating both sides with respect to $f$,  we derive
\begin{equation}
\der{}{f}\bra  x(\delta t)-x(0) \ket =\frac{D}{T} \delta t
-\frac{G}{2\gamma T} \delta t+O(\delta t^2)
\end{equation}
for large $\delta t$. This leads to the relation
\begin{equation}
\mud=\frac{2\gamma D-G}{2\gamma T}.
\label{mud-rel}
\end{equation}
{}Comparing  (\ref{final}) and (\ref{mud-rel}), we have arrived at 
(\ref{gamA}).

%%%%%%%%%%%%%%%%%%%%%%%%%%
%%%%%%%%%%%%%%%%%%%%%%%%%%
\paragraph*{Inequalities:}
%%%%%%%%%%%%%%%%%%%%%%%%%%
%%%%%%%%%%%%%%%%%%%%%%%%%%

%%  inequality 
As a simple application of the decomposition condition 
(\ref{hcon}), we derive several useful inequalities.  {}From the  
square of both sides of (\ref{Noise}) and the condition (\ref{hcon}), 
we obtain 
\begin{equation}
2 \Gamma^2 D=\delta t \bra (\delta B_n)^2 \ket +2 \gamma T,
\end{equation}
where we have used (\ref{noise:eff}). This immediately leads to an
inequality relating the intensity of the force noise in the original
system and the quantity representing its effective value in the 
coarse-grained system: 
\begin{equation}
\Gamma^2 D \ge \gamma T. 
\label{ine1}
\end{equation}
Because $\Gamma=\mud^{-1}$ (see (\ref{Gamma})), (\ref{ine1}) 
can be written as 
\begin{equation}
D \ge  \gamma \mud^2 T.
\label{ine2}
\end{equation}
{}We believe  that this  inequality holds in  other Brownian motors  
\cite{Butt,fla,pul} 
because the path integral expression is valid even for cases 
with a time dependent potential and the expressions  
(\ref{con2})-(\ref{mud-rel}) given in the proof are the same for 
those models. 
The inequality (\ref{ine2})  involves only  directly measurable 
quantities and therefore can be tested experimentally. 

%% remark on inequality 

In a related work, Sasaki 
conjectured that the inequality $D \ge \mud T$ holds generally 
for Brownian motors \cite{SSK}.  If we define  
the effective temperature $T_{\rm eff}$ using the FDR violation 
factor \cite{HSIIIa}, this conjecture is equivalent to the assertion 
that  $T_{\rm eff}$ is not less than the temperature of 
the environment, $T$. Because  it has been observed that 
$T_{\rm eff} > T$ in glassy systems \cite{glass2} and driven 
many-body systems \cite{drive1,HSIV}, the inequality 
$D \ge \mud T$ does seem plausible. However, Sasaki reported that 
this inequality is violated for the model (\ref{model1}) with  
an appropriate choice of $U(x)$ \cite{SSK2}.  This result leads us 
to believe that, in the present context, if there exists a generally
valid inequality among measurable quantities, perhaps it involves
the intensity of the force noise and its effective one, not the 
temperature and its effective one.

%%%%%%%%%%%%%%%%%%%%%%%%
%%%%%%%%%%%%%%%%%%%%%%%%
\paragraph*{Energetics:}
%%%%%%%%%%%%%%%%%%%%%%%%
%%%%%%%%%%%%%%%%%%%%%%%%

%% introduction

In equilibrium systems, heat is distinguished from work according 
to the second law of thermodynamics. However, obviously, heat 
can be considered as a mechanical work done by a force at a 
microscopic scale. This tempts us to investigate how heat and 
work come to be expressed in different ways through a coarse-graining 
procedure.  We treat this problem  on the basis of the decomposition 
of the force $-\overline{dU/dx}_n$ given by (\ref{decom}). 
    
%%  basics of energetics and question

In Langevin systems, the heat absorbed from a heat bath is 
interpreted as the work done by a force  $-\gamma dx/dt+\xi$ 
exerted by the heat bath  \cite{KS}. With this interpretation, 
the heat absorbed during an interval $t_n\le t\le t_{n+1}$ can  
be expressed as       
\begin{equation}
q_n=-\int_{t_n}^{t_{n+1}} \left(\gamma \frac{dx}{dt}-\xi\right)
\circ dx(t),  
\label{hm}
\end{equation}
where the symbol $\circ$ indicates that the integral here is 
the stochastic Stieltjes integral in the Stratonovich sense  \cite{KS}. 
Then, the energy balance equation for the model (\ref{model1}) is 
derived as
\begin{equation} 
U(x_{n+1})-U(x_n)=q_n+f(x_{n+1}-x_n).  
\label{b}
\end{equation}
Through similar considerations applied to the effective model 
(\ref{model2}),  we define the heat (absorbed from an ``effective 
heat bath'') as 
\begin{equation}
Q_n=-\left(\Gamma\frac{x_{n+1}-x_n}{\delta t}-\Xi\right)(x_{n+1}-x_n).
\end{equation}
With this, the energy balance equation for (\ref{model2}) is 
obtained as 
\begin{equation}
Q_n+F(x_{n+1}-x_n)=0. 
\label{mb}
\end{equation}  
The difference between the two quantities $q_n$ and $Q_n$ becomes 
obvious when their steady state averages are compared; we have 
$\bra q_n \ket=-f v_s\delta t$ and $\bra Q_n \ket=-F v_s\delta t$, 
while it is known that outside the linear response regime, in 
general $F\ne f$ (see Fig. 2 of  Ref. \cite{HSIIIa}). 

%% interpretation 

In order to understand the difference between $q_n$ and $Q_n$, we 
consider a decomposition of the work done by the force $-dU/dx$  
during a time interval $t_n\le t\le t_{n+1}$, which is written  as
\begin{equation}
-\overline{\frac{d U}{dx}\frac{dx}{dt}}_n\delta t  = \Qp_n + \Wp_n.  
\label{enerdec}
\end{equation}
Although we conjecture that $\Qp_n$ and $\Wp_n$ correspond to ``heat'' 
and ``work'',  respectively, no rule is known that distinguishes 
``heat'' from ``work'' in this case. However, in the present system, 
it seems natural to assume
\begin{equation}
\Wp_n=\fp (x_{n+1}-x_n),
\label{wh}
\end{equation}
because the extra driving force $\fp$ was determined from the 
decomposition condition (\ref{hcon}). With this assumption, we 
can derive the relation
\begin{equation}
Q_n=q_n+\Qp_n,
\label{addi}
\end{equation}
which provides a clear interpretation of the difference between 
$q_n$ and $Q_n$. 

%% perspective

The argument above leads to the following question: Is there a 
simple rule of energetics from which we can obtain the decomposition 
(\ref{enerdec}) along with (\ref{wh}) without the decomposition 
condition (\ref{hcon})?  This fundamental question 
will be studied in the future. In addition to their role in such 
fundamental problems, energetic considerations applied to different 
time scales may be useful when we attempt to interpret the efficiency 
of motor proteins \cite{YNG} within stochastic models \cite{TH}.  

%%%%%%%%%%%%%%%%%%%%%%%%%%%%%%
%%%%%%%%%%%%%%%%%%%%%%%%%%%%%%
\paragraph*{Adjoint dynamics:}
%%%%%%%%%%%%%%%%%%%%%%%%%%%%%%
%%%%%%%%%%%%%%%%%%%%%%%%%%%%%%

% definition of adjoint dynamics

As a final topic here, we consider the relation between the 
decomposition condition (\ref{hcon}) and  time reversal symmetry.  
In order to represent this symmetry explicitly, we consider the 
path probability density $P$ for a discrete time series $[x]_N=
(x_0,x_1,\cdots,x_N)$  generated within the  model under 
consideration. The time-reversed path probability density $P^*$ is 
defined by $P^*([x]_N)=P(\tilde{[x]}_N)$, where  $\tilde{[x]}_N$ 
represents the time reversed trajectory of $[x]_N$, that is, 
$\tilde x_{n}=x_{N-n}$. When $P^*$ is obtained  from the frequency 
distribution of trajectories in a steady state for some stochastic 
dynamics, these dynamics are called ``adjoint dynamics''. 

% result and proof

When the constant $A$ appearing in (\ref{decom}) is chosen correctly, 
(\ref{model3}) can be regarded as an effective Langevin model.  
In this  case, we find that  adjoint dynamics are described by    
\begin{equation}
(\gamma-A) \frac{x_{n+1}-x_n}{\delta t}= -f- \fp +\delta B_n+ 
\overline{\xi}_n.
\label{model42}
\end{equation}
Then, note that  the condition (\ref{hcon}) is equivalent to 
\begin{equation}
\bra(-\delta B_n+\overline\xi_n)^2  \ket
=\bra (\delta B_n+\overline\xi_n)^2 \ket.
\label{con}
\end{equation}
Because both $\delta B_n$ and $\bar \xi_n$ exhibit Gaussian 
distributions for sufficiently large $\delta t$, the condition 
(\ref{con}) allows us to replace $\delta B_n+ \bar \xi_n$ in 
(\ref{model42}) by $-\delta B_n+ \bar \xi_n$. Thus, the adjoint 
dynamics can be expressed as
\begin{equation}
(\gamma-A) \frac{x_{n+1}-x_n}{\delta t} = -f-B_n+\overline{\xi}_n.
\label{model41} 
\end{equation}
Then, because (\ref{model3}) can be  rewritten as 
\begin{equation}
(\gamma-A) \frac{x_{n+1}-x_n}{\delta t} = f+B_n+\overline{\xi}_n, 
\label{model50} 
\end{equation}  
we find that the decomposition of $-\overline{dU / dx}_n$ given by 
(\ref{decom}) is characterized by parity with respect to time reversal  
in the stochastic sense. More specifically, the dissipative force 
$A (x_{n+1}-x_n)/\delta t $ remains, while the other contribution,  
$B_n$,  changes sign in the adjoint dynamics.  

%% Bertini  et. al. %% 

In a previous study related to adjoint dynamics, Bertini et al.  
succeeded in deriving the large deviation functional $S$ of the 
density profile for a special model of the hydrodynamic equation 
$\partial_t\rho=\cD(\rho)$, which is obtained as the continuum 
limit for a nonequilibrium lattice gas \cite{Ber}. In their analysis, 
the equation describing the adjoint dynamics $\partial_t\rho=
\cD^*(\rho)$ was rigorously derived in the form  
\begin{equation}
\cD^*(\rho) = \frac{1}{2}\nabla\left(\chi(\rho)\nabla
\frac{\delta S}{\delta \rho} \right )-\cA(\rho)
\label{ber2}
\end{equation}
for the case in which $\cD(\rho)$ is given by 
\begin{equation}
\cD(\rho) = \frac{1}{2}\nabla\left(\chi(\rho)\nabla
\frac{\delta S}{\delta \rho}\right) +\cA(\rho), 
\label{ber}
\end{equation}
where $\chi(\rho)$ is the current noise intensity. They called the 
relation $\cD(\rho)+\cD^*(\rho)=\nabla\left(\chi(\rho)\nabla
(\delta S/\delta \rho)\right)$  the ``fluctuation dissipation 
relationship''  for NESSs far from equilibrium, because it reduces 
to one expression among the linear response relations for states near  
equilibrium, where the relation  $\cD(\rho)=\cD^*(\rho)$ holds due 
to detailed balance. We remark that  (\ref{ber2}) and  (\ref{ber}) 
are similar to  (\ref{model41}) and (\ref{model50}) in the model we 
study.

%%% many-body systems %%%

%%%%%%%%%%%%%%%%%%%%%%%%%%%%%%
%%%%%%%%%%%%%%%%%%%%%%%%%%%%%%
\paragraph*{Conclusion:}
%%%%%%%%%%%%%%%%%%%%%%%%%%%%%%
%%%%%%%%%%%%%%%%%%%%%%%%%%%%%%

In conclusion, we have found that the decomposition condition 
(\ref{hcon}) leads to the relation  $\Gamma=\mud^{-1}$, which was 
obtained in our previous study \cite{HSIIIa}.  The condition 
(\ref{hcon}), which is related  to time reversal symmetry, yields 
the new inequality (\ref{ine2}) and leads to an interesting question 
regarding the decomposition of work given in (\ref{enerdec}).

One of the authors (K. H.) acknowledges her mother for encouragement 
over the past decade. This work was supported by a grant from the 
Ministry of Education, Science, Sports and Culture of Japan 
(No. 16540337).

\end{document}